\documentclass[12pt]{article}

\begin{document}

\title{Universality of vortex avalanches in a type II superconductor with
periodic pinning}

\author{R. Cruz\\Natural Science Faculty, ISPH Holgu\'{\i}n, Cuba\\
Superconductivity Laboratory, IMRE-Physics Faculty,\\
University of Havana\\10400 La Habana, Cuba \\ 
\\
R. Mulet and E. Altshuler\\Superconductivity Laboratory, IMRE-Physics
Faculty,\\
University of Havana\\10400 La Habana, Cuba \\}
\date{\today}

\maketitle

\begin{abstract}

In this work the robustness of a simple cellular automaton
developed by Bassler and Paczuski\cite{Bassler}
to describe the critical state in type-II
superconductors is studied. Two different configurations of 
pinning centers are introduced
and a new universality class is found. 
The
numerical values of the critical exponents were calculated following two
scaling techniques to ensure the validity of our results.

\end{abstract}

\section{Introduction}
\label{sec_introduction}

In type-II superconductors, if the external magnetic field is greater than
the first critical field $H_{c1}$, vortices penetrate the material. 
These vortices move freely in ideal superconductors if an electrical current 
is applied, destroying superconductivity \cite{Tinkham}. However, in 
real materials, the existence of pinning centers 
prevent vortex 
motion if the current is less than a certain value, called critical 
current.

Bean in 1966 \cite{Bean} proposed that the distribution of magnetic field
inside a real sample, when the external magnetic
field increses over $H_{C1}$, can be represented by a linear profile in a
$H(x)$ graph, where the slope corresponds to the critical current density.

In 1989 Bak et al \cite{Bak}, in an attempt
to explain the behavior of many
dynamical systems, developed
 the concept of self organized criticality (SOC).
In their approach, after long time of avalanches evolution, those systems reach 
a complex state
which exhibits a power law in their avalanche size distributions and in their
avalanche duration distributions, and $f^{-\beta}$ noise.
Sandpiles have become a paradigm of these sytems, and their closed similitude
with the critical state of superconductors \cite{de_gennes} 
quickly called the
attention of researchers in the field \cite{Tang} \cite{Mulet}
\cite{Vinokur}.
 
In fact, Field et al in 1996 \cite{Field} 
experimentally observed SOC in vortex avalanches at the inner wall of a 
low-$T_{c}$ superconducting hollow cylinder submitted to slowly ramped
axial magnetic fields. In a similar experiment performed on a low $T_{c}$
thin film ring, Nowak et al \cite{Nowak} observed either potential or
peaked distributions of vortex avalanche size,  depending on temperature.
Finally, in a recent experiment, Behnia et al \cite{Maiily} 
measured the internal vortex avalanches in a low-$T_{c}$ film submitted to
an increasing magnetic field, and found power distributions of avalanche
size only for temperatures higher than $3K$.

On the other hand, many simulations had been performed on 
vortex avalanches.
In 1994, Richardson et al \cite{Pla} developed a molecular dynamic model 
showing the discrete evolution of the magnetic profiles inside the
superconductors. Also using
molecular dynamic techniques, Olson et al \cite{Olson} tried to mimic 
Field's experiment.  They found power distributions of avalanche sizes for
high densities of pinning centers.
However, none of these simulations was able to correctly estimate the
critical exponents involved in the SOC theory for this system because of
strongly finite size effects.

More recently, Bassler and Paczuski \cite{Bassler} proposed a
simple cellular automata to model the dynamics of magnetic flux motion at
a length scale larger than the range of vortex interaction
for a system with a random distribution of pinning centers. They obtained, using
finite size scaling techniques, the existence of critical exponents for a
wide distribution of parameters in their simulation, and hence, of SOC.
In this work, we explore the robustness of this cellular automaton 
varying the distribution and depth of the pinning centers. 

In the next section we explain
Bassler and Paczuski's model.In section 3 we present and discuss 
the results obtained using two novel pinning
centers configurations: periodic distributions and combined periodic plus
random distributions. Finally the conclusions are given.

\section{Model}
\label{sec_model}

 Bassler and  Paczuski's cellular automaton\cite{Bassler} is a two dimensional hexagonal lattice where each
site is occupied by $m(x)$ vortices. Vortices in the site 
$x$ can move towards  the site $y$ only if the force acting on them, 
in that direction, is greater than zero. 
This force is calculated using the following formula:

\begin{eqnarray}
F=V(y)-V(x)+(m(x)-m(y)-1)+ \nonumber \\
	r(m(x_{1})+m(x_{2})-m(y_{1})-m(y_{2}))   
		\label{eq:Fuerza}
\end{eqnarray}
      
\noindent where $x_{1}$, $x_{2}$ and $y_{1}$, $y_{2}$ are the other nearest neighbors
of $x$ and $y$, respectively, and $V(x)$ and $V(y)$ are the 
strengths  of the pininng centers at those sites. 
If the distributions of pinning centers is
random, $V(x)=p$, with a probability $q$, and $V(x)=0$ with a 
probability $1-q$. The parameter $r$ (where $r<1$)
characterizes the long distance action of the next nearest neighbours.
If there is more than one unstable direction, one of them is chosen at random. All
lattice site are updated in parallel, and at each site only one vortex can
move on a particular update.
Periodic boundary conditions are applied to the top and the bottom of the
lattice. The vortices that reach the right edge of the system are removed, 
while they are not allowed to abandon the system through its 
left edge.

An avalanche begins by randomly choosing a site at the left edge of the system
and adding one vortex to it.
It continues with the consecutive
update of the lattice sites until no more unstable sites persist. 
Once the lattice is again stable, another vortex is added. 
The avalanche size is defined as the number of toplings corresponding to the
addition of one vortex while the avalanche duration is defined as the number
of updatings necessary to complete one avalanche.

To characterize the system, in analogy with other SOC models, Bassler 
 and Paczuski\cite{Bassler} proposed
and proved the following scaling ansatz for the probability distribution of
avalanches sizes~(\ref{eq:DistA}), and avalanches
durations~(\ref{eq:DistT}):

\begin{equation}
P(s,L)=s^{-\tau} g(s/L^D)                  \label{eq:DistA}
\end {equation}                                                                        

 \begin{equation}
P(t,L)=t^{-{\tau}_t} g(t/L^z)              \label{eq:DistT}
\end {equation}                                                                        

\noindent and obtained the following set of scaling exponents 
for a wide range of
parameters values in the model $\tau = 1.63\pm0.02$, 
$D = 2.7\pm0.1$, $\tau_{t} = 2.13\pm0.08$
and $z = 1.5\pm0.1$
which are related by the scaling relations:

\begin{equation}
	\tau(2D-1) = 1	                      \label{eq:SC01}	
\end{equation}
\begin{equation}
	D(\tau-1) = z(\tau_t-1)			\label{eq:SC02}
\end{equation}

In our work, to ensure the validity of our results, we also used the following
 scaling ansatzs \cite{A.DIAZ} for the avalanches
sizes and avalanches durations, respectively: 

\begin{equation}
P(s,L) = L^{-\beta} f(s/L^D)                  \label{eq:DistA1}
\end {equation}                                                                        

 \begin{equation}
P(t,L) = L^{-\omega} f(t/L^z)                 \label{eq:DistT1}
\end {equation}                                                                        

Therefore two new scaling exponents characterize our system $\beta$ and $\omega$.
They are related to the previous ones by: $\beta=\tau D$ and $\omega=\tau_{t}
z$, and satisfy the following scaling relations:

\begin{equation} 
\beta = 2D-1                               \label{eq:ESC1}        
\end{equation} 

\noindent and

\begin{equation} 
\omega = \beta+z-D                           \label{eq:ESC2}    
\end{equation}

\noindent With this notation,  Bassler and Paczuski's results \cite{Bassler} 
give $\beta = 4.4$, $D = 2.7$, $\omega = 3.19$ and $z = 1.5$.  

\section{Results and Discussion}
\label{sec_results}

Following Bassler and 
Paczuski\cite{Bassler}, we first assumed a random distribution of
pinning centers, with parameters $(r,p,q)=(0.1;3;0.1)$. 
In figure 1
appears  the collapse of our results for dimensions 
$L=100,160,200$ 
using the scaling ansatz~(\ref{eq:DistA1}) and ~(\ref{eq:DistT1}). 
The critical exponents for these conditions are in good agreement with
reference \cite{Bassler}. Similar results were obtained for other parameters within
the range already checked in \cite{Bassler}, in all case after a $10^7$m.c.s . 

Then, we tested the model using a combination of periodic and random 
pinning distributions. This kind of configuration introduces 
a new parameter 
to geometrically
characterize the system i.e, the distance between the periodic pinning
centers $a$. In our simulations we used  
$r=0.1$, and a constant density of random pinning centers $q=0.1$. 
Like real irradiated superconducting materials, 
we used different pinning strenghts $p_{1}$ and $p_{2}$ for periodic and
random pinning centers, respectively.
The periodic pinning centers were always chosen stronger or equal to the
random ones trying to mimic irradiation effects.

For the sets of parameters $(p_{1};p_{2};a)= 
(10;1;20)$, $(10;1;10)$, $(5;1;10)$, $(10;1;4)$ and $(20;1;4)$ the 
critical exponents conserved the previously reported values of: 
$\tau= 1.63\pm0.04$ , $\tau_t = 2.13\pm0.09$, $\beta= 4.4\pm0.1 $, 
$D = 2.7\pm0.1$, $\omega = 3.2\pm0.2  $ and
$z=1.5\pm0.1$ 
for different system dimensions, indicating the robustness of the system, 
(see the two data collapse
presented in figure 2). 

We also performed simulations using only a periodic
pinning configuration. It was demostrated
that the system also conserved    
the critical exponents for the sets of parameters $(p;a)=(5;10)$
$(5;20)$, $(5;4)$, $(1;10)$ as is shown in figure 3. 
However, as the data collapse in figure 4 displays, 
using lower values of $a$ and stronger pinning 
centers (for example: 
$(10,4)$ and $(20,4)$) caused a decreased in
the exponents $\beta, \tau, \tau_t$ and $\omega$  
to $\beta = 3.2\pm0.1$, $\tau = 1.45\pm0.02$, 
$\tau{_t} = 1.7\pm0.08$ and $\omega = 2.6\pm0.2$, while
 $z = 1.6\pm0.1$ and $D = 2.2\pm0.1$, in good
agreement with the scaling
relations~(\ref{eq:SC01})~(\ref{eq:SC02})~(\ref{eq:ESC1})
and~(\ref{eq:ESC2}).This
suggests the existence of a new universality class 
for dense and strong periodic pinning configurations.

To study  the origin of this new universality class we made
calculation using a superposition of random 
and periodic pinning configurations with equal 
potentials $p{_1}=p{_2}$. The scalings results, presented 
in the figure 5, show 
that systems with parameters $(p{_1};p{_2};a)=(10;10;4)$ 
conserved Bassler and Paczuski's\cite{Bassler} exponents values.

These 
results and those
obtained using only a periodic configuration with the set of parameters 
$(p;a)=(5;4)$  (see figure 3) proved  
that the new universality class appears due to the joint action of the
strong periodic pinning centers and the neglegible influence of the 
random pinning distribution. If in the system one of these conditions is 
absent,it will have a "normal" evolution.            

The existence of the new universality class could be explained 
based on the following qualitative argument.
The presence of strong and correlated pinning, the neglegible influence of
the random pinning and 
the low values of $a$ could produce "magnetic traps" \cite{Olson}
were vortices
oscillate between neigbour pinning centers, increasing the number 
of bigger avalanches, and then reducing the
characteristic exponents, $\tau, \tau_t, \beta$ and $\omega$.

Finally an experiment is suggested to check the existence of this new
universality class. It consist in performing an avalanche detection
experiment similar to the one reported in reference \cite{Maiily}, but before an
after irradiating the sample with heavy ions in order to obtain arrays of
periodic "strong" pinning centers as reported by Harada et al \cite{Harada}.
If some relevant experimental parameters such as the magnetic field sweep rate
and the temperature are conveniently tuned, critical exponents close to those
reported in \cite{Bassler} should be achived before irradiation. After the
irradiation (under the same set of parameters) the combination of "weak"
random pinning plus "strong" periodic pinning is expected to move the
critical exponents towards the values reported here.

\section{Conclusions}
\label{sec_conclusions}
A cellular automaton model for vortex avalanches was analized in the presence
of periodic-random and periodic distributions of pinning centers.
The robustness of the model was proved
for a wide range of parameters and
the existence of a new universality class for strong and dense
periodic distribution of pinning centers was shown. 
A qualitative argument to support the reasons
for the 
appareance of this new
universality class were given.

\section{Acknowledgments}
We thank A.J. Batista-Leyva for helpful discussions and D.Bueno and M.Miriela
for their collaboration during the calculations. The computing facilities
donated by S. Garc\'{\i}a are highly appreciated. We also acknowledge the
"Alma Mater Project" 1998 for financial support.

\newpage

\section*{Figure captions}
{\bf Figure 1} Finite scaling plot using~(\ref{eq:DistA1}) 
$L=100,160,200$ and $(r;p;q)=(0.1;3;0.1)$. 
Inset: finite scaling plot using~(\ref{eq:DistT1})    
\newline
{\bf Figure 2} Finite scaling plot using~(\ref{eq:DistT1}) 
$L=100,160,200$ and
$(r;p_{1};p_{2};a)=(0.1;10;1;10)$. Inset: finite scaling 
plot using~(\ref{eq:DistA}) 
\newline
{\bf Figure 3} Finite scaling plot using~(\ref{eq:DistA1}) 
$ L-100,160,200$ and $(p;a)=(5;4)$. 
Inset: finite scaling plot using~(\ref{eq:DistT}) 
\newline
{\bf Figure 4} Finite scaling plot using~(\ref{eq:DistT1}) 
$L=100,160,200$ and $(p;a)=(10;4)$. 
Inset: finite scaling plot using~(\ref{eq:DistA}) of 
\newline
{\bf Figure 5} Finite scaling plot using~(\ref{eq:DistT1}) 
$L=100,160,200$ and $(r;p_{1};p_{2};a)=(0.1;10;10;4)$. 
Inset: finite scaling plot using~(\ref{eq:DistA}) 

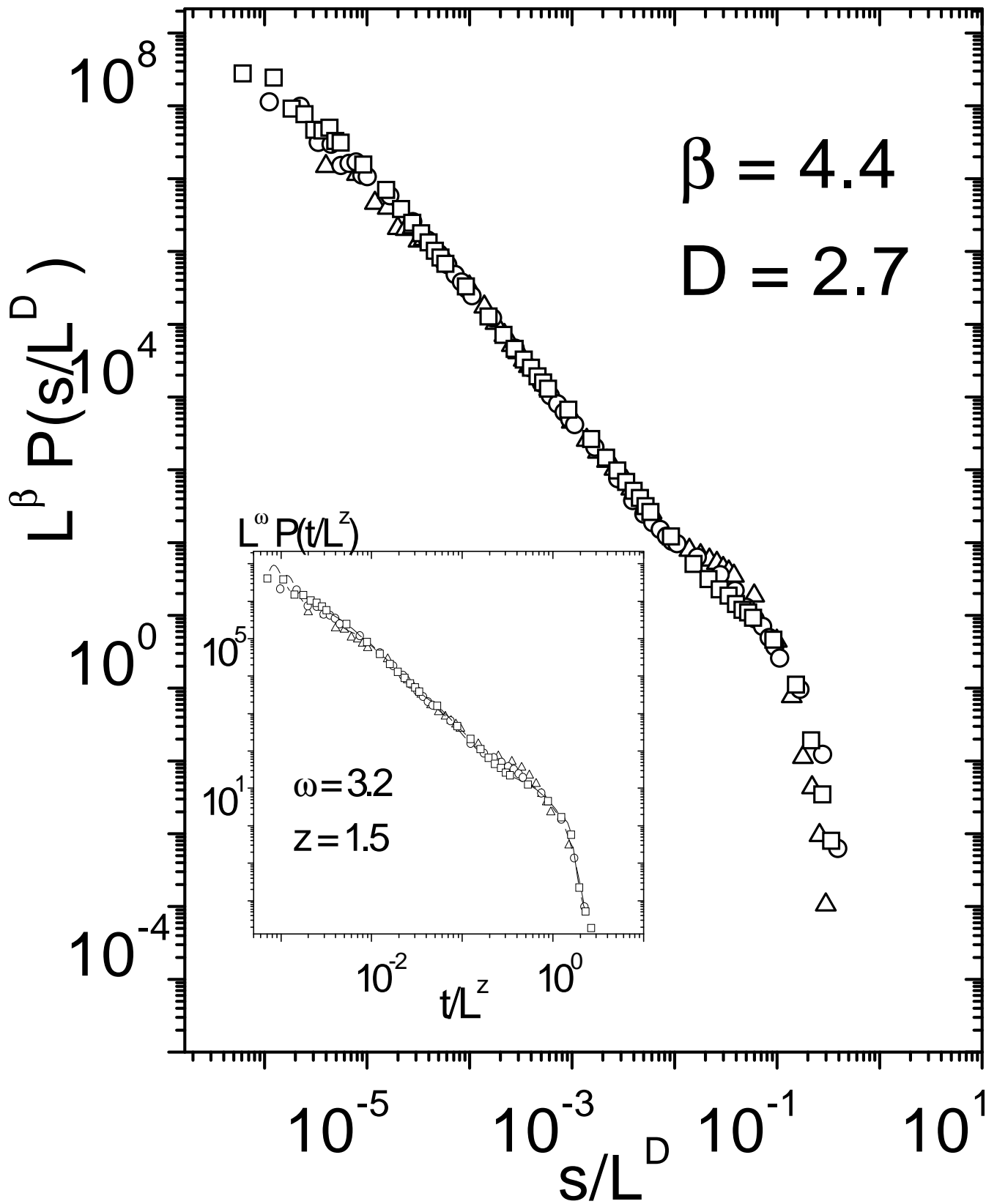
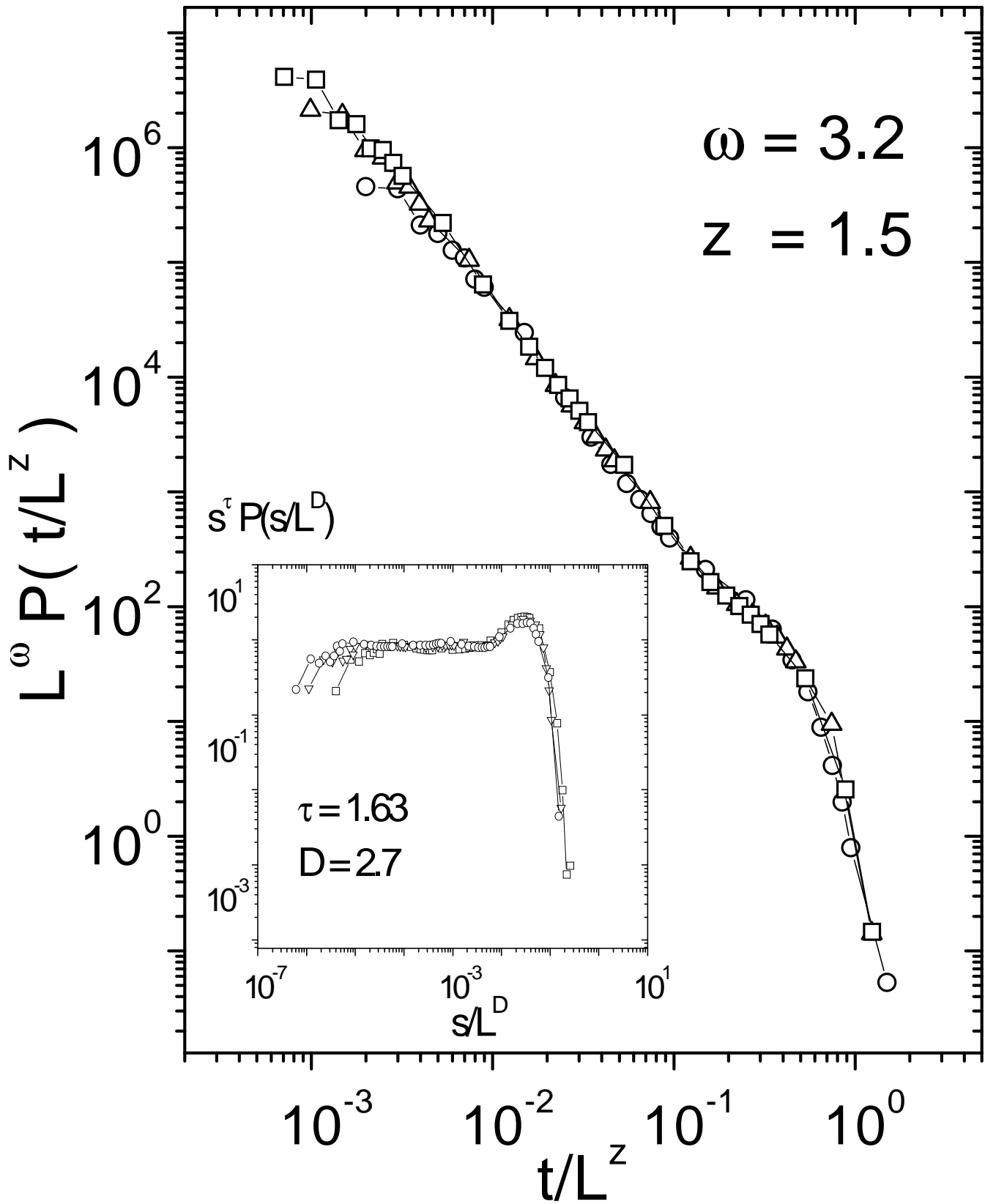
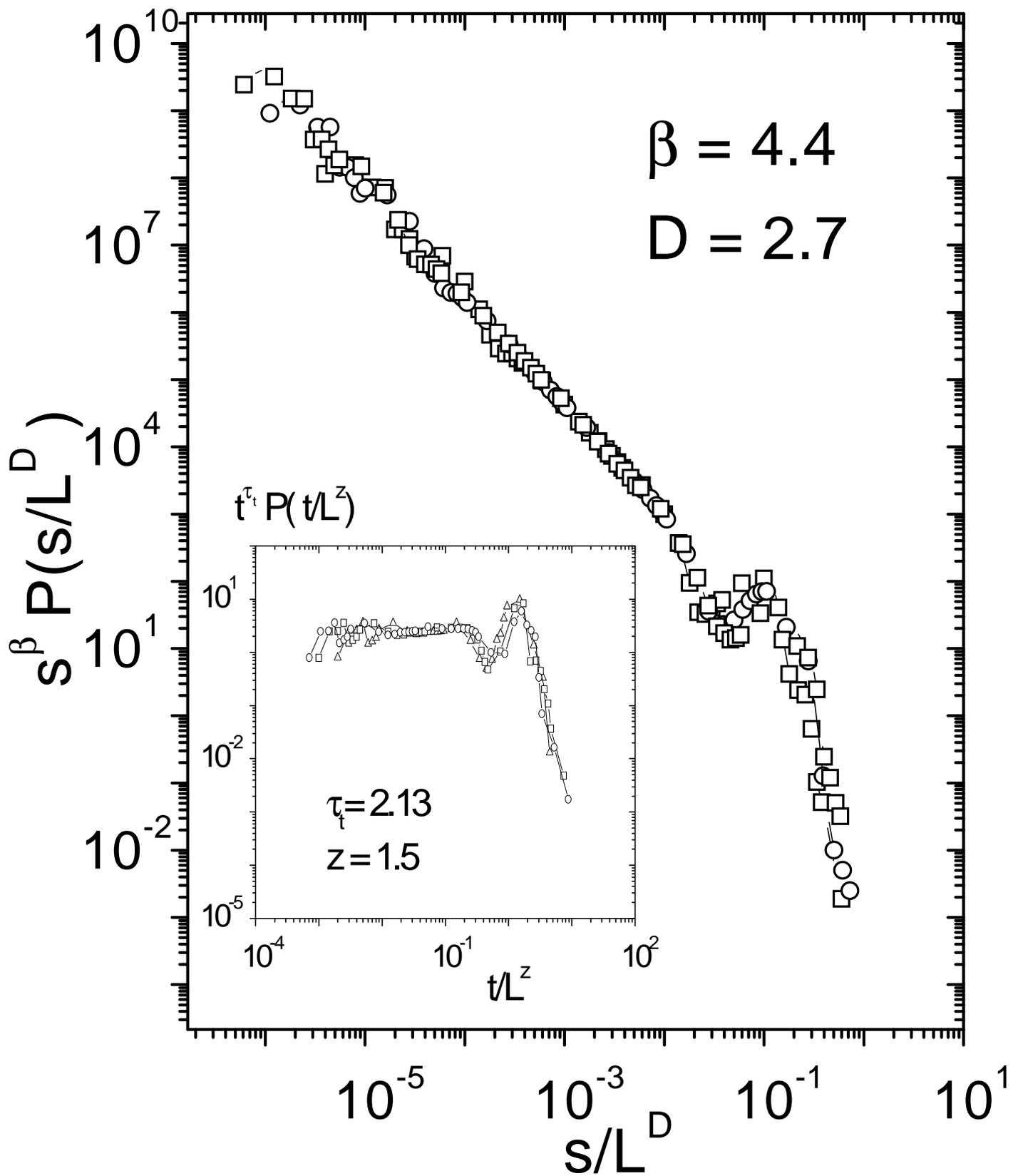
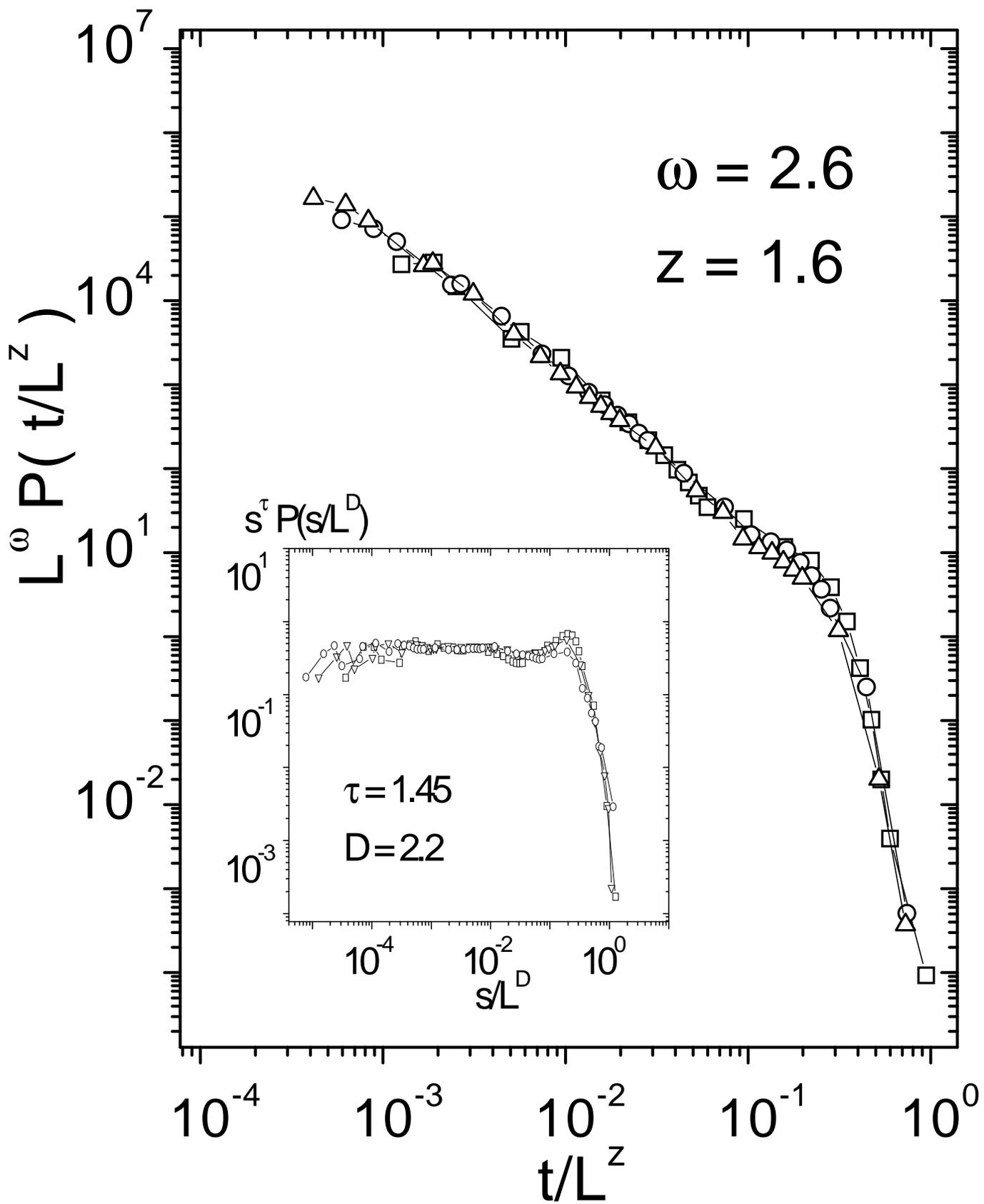
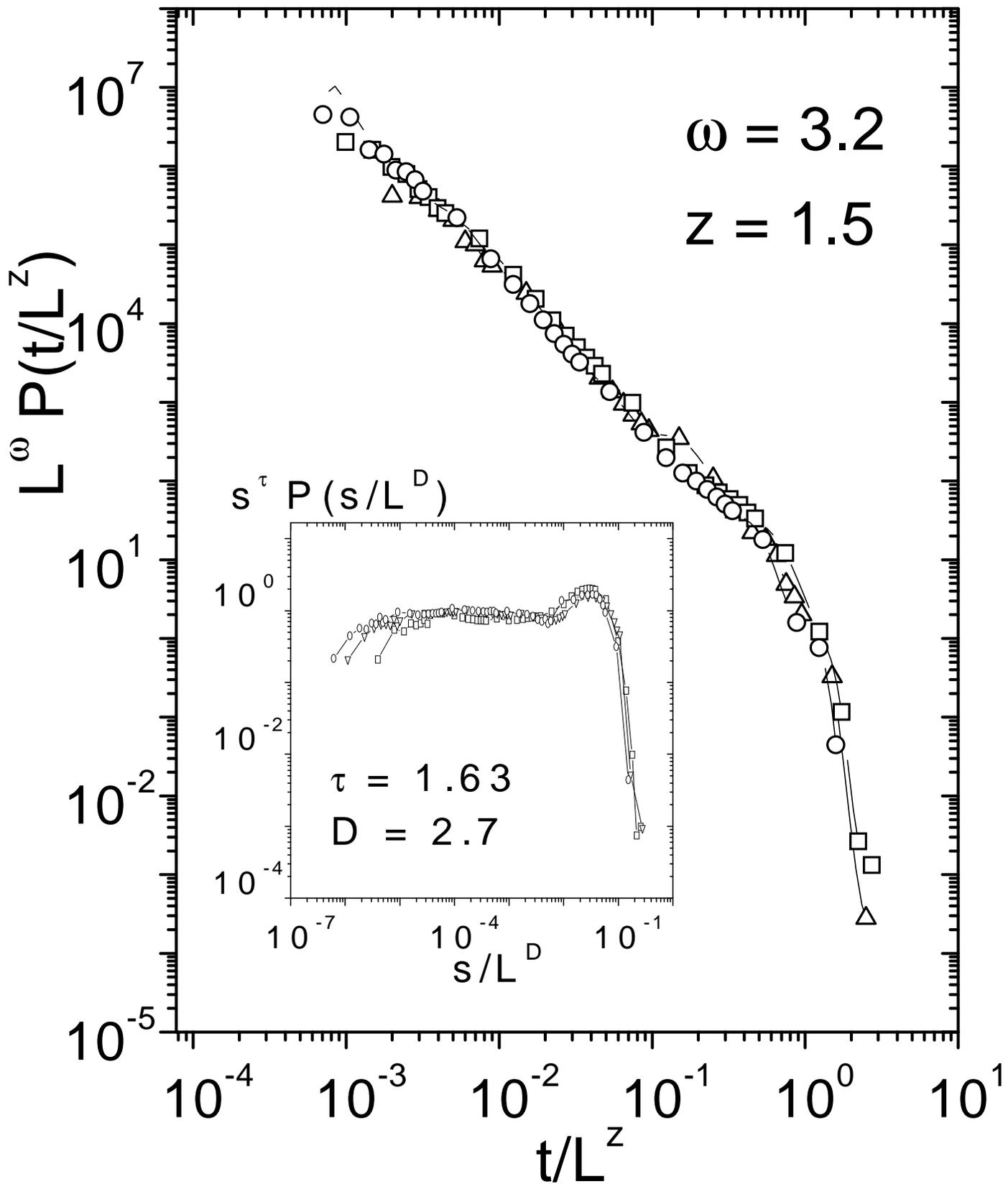
\end{document}